\documentclass[preprint,aip,jcp]{revtex4-1}
\usepackage{graphicx}
\usepackage{bm}
\usepackage{amsmath}
\DeclareMathOperator{\sgn}{sgn}
\usepackage{amssymb}
\usepackage{subfigure}
\usepackage{gensymb}
\usepackage{epstopdf}
\usepackage{color}

\usepackage[normalem]{ulem}
\usepackage{csquotes}
\begin{document}
\title{Spherically symmetric solvent is sufficient to explain lower critical solution temperature in polymer solutions}
\author{G. Swaminath Bharadwaj}
 \affiliation{Department of Chemical Engineering, Indian Institute of Technology Madras, Chennai-600036, India.}
\author{P. B. Sunil Kumar}
 \email{sunil@iitm.ac.in}
\affiliation{ 
Department of Physics, Indian Institute of Technology Madras, Chennai-600036, India}
\author{Shigeyuki Komura}
 \affiliation{Department of Chemistry, Graduate School of Science and Engineering, Tokyo Metropolitan University, Tokyo 192-0397, Japan.}
\author{Abhijit P. Deshpande}
 \affiliation{Department of Chemical Engineering, Indian Institute of Technology Madras, Chennai-600036, India.}
\date{\today}
\begin{abstract}
We study the lower critical solution temperature  (LCST) in thermoresponsive polymer solutions by means of a coarse-grained single polymer chain simulation and a theoretical approach. The simulation model includes solvent explicitly and thus accounts for  solvent interactions and entropy directly. The theoretical model consists of a single chain polymer in an implicit solvent where the effect of solvent is included through the intra-polymer solvophobic potential proposed by Kolomeisky and Widom. Our results indicate that the LCST behavior is determined by the competition between the mean energy difference between the bulk and bound solvent, and the entropy loss due to  the bound solvent. At low temperatures, solvent molecules are bound to the polymer and the solvophobicity of the polymer is screened, resulting in a coiled state.  At high temperatures the entropy loss due to bound solvent offsets the energy gain due to binding which causes the solvent molecules to unbind, leading to the collapse of the polymer chain to a globular state. Furthermore, the coarse-grained nature of these models indicates that mean interaction energies are sufficient to explain LCST  in comparison to specific solvent structural arrangements.
\end{abstract}
\maketitle

\section{Introduction}
Thermoresponsive polymers are an important class of materials which exhibit temperature dependent structural changes and find application in  drug delivery,\cite{Hoare2009,Okano1995a} surface modification,\cite{Hoare2008} and self-assembled structures.\cite{Sun2013} Our interest lies in the family of thermoresponsive polymers which exhibit a lower critical solution temperature (LCST) in aqueous solutions. Along with the LCST, these polymers also exhibit a coil-to-globule transition at the single chain level. A well known example of such thermoresponsive polymers is Poly(N-isopropylacrylamide) (PNiPAM) which exhibits a  LCST  in water and a upper critical solution temperature (UCST) in solvents such as methanol, ethanol, dimethyl sulfoxide, acetone.\cite{Costa2002a,  Bischofberger2014a} There have been several experimental \cite{Wang1998, Wu1998, Ono2006} and simulation \cite{Deshmukh2009a, Deshmukh2011, Alaghemandi2012a, Deshmukh2012a, Tucker2012a, Chiessi2010a} studies on the mechanism of the LCST behavior of PNiPAM in water.  The tunability  of the LCST for different applications has also been explored by studying its variation with additives such as salt,\cite{Zhang2005a, Du2010a} surfactant,\cite{Schild1991a, Shinde2001a, Mohan2007a} co-solvents,\cite{Zhang2002, Scherzinger2014, Mukherji2013, Mukherji2014, Mukherji2015, Walter2010a} and by the change in macromolecular architecture such as branching and tacticity. \cite{Ray2005a, Katsumoto2008a}

The origin of LCST in thermoresponsive polymer solutions is an important question in the field of polymer science. To understand this phase transition, there have been several attempts,  ranging from the mean-field theory to atomistic molecular dynamic simulations,\cite{Deshmukh2009a, Du2010a, Deshmukh2011, Alaghemandi2012a, Deshmukh2012a}  focusing  on the LCST of PNiPAM-water system. A mean-field model was proposed by Okada and Tanka \cite{Okada2005} who hypothesized that preferential interaction among bound water molecules (cooperative hydration) controls the transition. The simulation results of Deshmukh \textit{et al.}\cite{Deshmukh2012a}  indicate that the stability of bound water structure is the driving force for the transition. Though insightful, one should keep in mind that the results of these simulations are  obtained for a particular polymer-solvent system, and these approaches require very specific interactions,  extensive chemical details, or forcefield parameters.  Schild \textit{et al.}\cite{Schild1991} have studied the LCST behavior of PNiPAM in water-alcohol mixtures using a combination of experiments and the three-component Flory-Huggins model. In their model, the interaction parameters have been partially taken from experimental data, and it does not give us a clear idea about the generic mechanisms that lead to this phenomena.\\
\indent In the experimental studies, a rich diversity of material systems based on PNiPAM have been explored to obtain physical insights related to the effect of substituents, copolymers, solvents, and additives. From the view point of the mechanism, Ono and Shikata\cite{Ono2006} have calculated the number of water molecules per monomer using high frequency dielectric relaxation measurements.  Their results showed  that the LCST is driven by the complete dehydration of the PNiPAM chains, showing the importance of bound water near the polymer. Bischofberger and coworkers \cite{Bischofberger2014, Bischofberger2014a} have performed turbidity and dynamic light scattering on the ternary system of PNiPAM, water, and alcohol. Their results indicate that the thermodynamic description of the solvent is more important than the specific description of local solvent structure.\\
\indent Given the multiplicity of systems that can exhibit LCST, it is highly pertinent to come up with a model that can exibit LCST broad physical principles.  An approach aimed at identifying the minimal model that exhibits LCST will help to understand the relative importance of different contributions.  Generic polymer models\cite{Anderson2006a, Marrink2007a, Mella2010a, Polson2005, Hatakeyama2007a} are suitable candidates for this kind of approach. The coarse-grained nature of these models allows us to qualitatively study the importance of  the competition between  entropy and internal energy without invoking to a specific polymer or solvent. In this paper, we develop generic polymer models with spherically symmetric solvent and monomeric beads for simulation and theoretical studies of a coil-to-globule transition. Our results indicate that the LCST depends on the competition between the mean interaction energy difference between the bulk and bound solvent, and entropy of bound solvent. We show that  a coarse-grained representation of the solvent is sufficient to exhibit a LCST behavior. This indicates that the mean interaction energy difference between the bound solvent and bulk solvent is more important in comparison to the structural arrangement of the bound solvent. An important point to note is that this work is aimed at a generic understanding of the LCST behavior in thermoresponsive polymers, that does not refer to any particular polymer.\\
\indent The rest of the paper is organized as follows: in Sec.~\ref{sec:model}, we propose a polymer-solvent model for molecular dynamic simulation studies and introduce the solvophobic potential\cite{Kolomeisky1999} used in the theoretical approach. Section ~\ref{sec:results} presents the simulation results and numerical calculations of the theoretical model. In Sec.~\ref{sec:sum}, our findings will be summarized.
\section{Models}
\label{sec:model}
 In this section, we discuss  our  models used for the simulation and theoretical  studies.  While simulations are carried out using a bead-spring model for polymers in an homogenous single component solvent,  a phenomenological model is used for  the theoretical analysis. Below we describe these models in detail.
\subsection{Generic polymer model with explicit solvent}
\begin{figure*}[h!]
\begin{center}    
\includegraphics[scale=0.22]{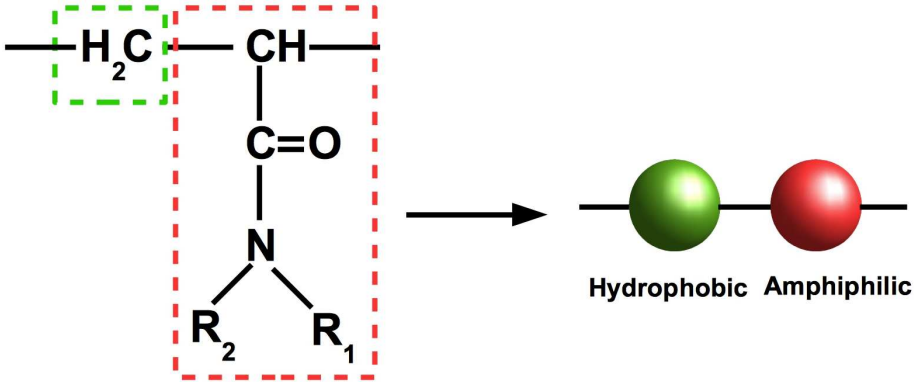}
\end{center}
\caption{Ad-hoc mapping of the acrylamide family of thermoresponsive polymers. $R_{\rm 1}$ and $R_{\rm 2}$ can be any arbitrary groups. The part of the monomer within the red box may have both hydrophilic and hydrophobic groups due to which it is modeled as an amphiphilic bead.}
\label{fig:mapping}
\end{figure*}

% Motivation for choosing the model: Link to the structure of poly(N,N alkyl alkyl acrylamide)
For the simulation studies, we model the polymer as a linear chain consisting of alternating solvophobic and amphiphilic beads ($N$ total beads, $N/2$ solvophobic, and $N/2$ amphiphilic beads). The motivation behind the presence of two different kinds of beads is to capture the behavior of the acrylamide family of thermoresponsive polymers in a generic manner (see Fig.~\ref{fig:mapping}). The methylene units along the backbone are analogous to hydrophobic beads. The substituted methylene units, most generally will have both hydrophilic and hydrophobic groups, and therefore analogous to amphiphilic beads. We emphasize that the intention is to use a generic model, without relating to any specific polymer system.

 The amphiphilic bead has attractive interactions with both the solvent and the solvophobic beads. The interaction between the solvophobic bead and the solvent is purely repulsive. The solvent can be introduced  either by including it explicitly or by incorporating its effects implicitly within the interaction potentials. Since coarse-grained potentials are obtained by integrating out the  internal degrees of freedom of the unit, it is temperature dependent in general. When the scale of coarse graining is small, such a dependence is weak and can be neglected.  However,  when the solvent  is  implicit, the interaction potentials   have a stronger dependence  on the temperature, and the nature of this dependence has to be assumed a priori. To avoid a specifically assumed temperature dependence of the interaction potentials, we explicitly 
incorporate the solvent. The potential energy for the system is given by the following expression
 \begin{widetext}
 \begin{eqnarray}\label{eq:energy_simu}
 E=\sum_{i=1}^{N-1} k_{\rm b}(b_{i}-b_{i0})^{2} + \sum_{i=1}^{N_{\rm t}}\sum_{j>i} 4\epsilon_{ij}\Bigg[\left(\frac{\sigma}{r_{ij}}\right)^{12}-\left(\frac{\sigma}{r_{ij}}\right)^{6} - \left(\frac{\sigma}{r_{{\rm c},ij}}\right)^{12}+\left(\frac{\sigma}{r_{{\rm c}, ij}}\right)^{6}\Bigg],
 \end{eqnarray}
 \end{widetext}
where $N$ is the number of  beads in the polymer chain, as mentioned before. $k_{\rm b}$ the force constant for the bonded interaction, $b_{i}$ the bond length between neighboring beads, $N_{\rm t}$  the total number of beads in the system (polymer + solvent), $b_{i0}$ the equilibrium bond length and $r_{ij}$ the distance between two non-bonded beads. The second term is the Shifted Lennard Jones (SLJ)  potential with $r_{{\rm c},ij}$ being the cutoff distance at which the potential is truncated and shifted to zero. The above form of SLJ potential ensures that all the beads are spherically symmetric  and have size  $\sigma$. All the interaction parameters are kept independent of the temperature. We define dimensionless quantities as  $\overline{r}_{ij}=r_{ij}/\sigma$, $\overline{\epsilon}_{ij}=\epsilon_{ij}/\epsilon_{\rm ss}$, $\overline{k}_{\rm b}=\sigma^{2}k_{\rm b}/\epsilon_{\rm ss}$, $\overline{b}_{i0}=b_{i0}/\sigma$, $\overline{T}=k_{\rm B}T/\epsilon_{\rm ss}$, $\overline{P}=\sigma^{3}P/\epsilon_{\rm ss}$ and $\overline{t}= t\sqrt{\epsilon_{\rm ss}/(m\sigma^2)}$, where $\epsilon_{\rm ss}$ is the potential energy of interaction between two solvent beads. We fix the values to $\overline{b}_{i0}=1$ and $\overline{k}_{\rm b}=200$ for all the simulations. The values of the other interaction parameters are listed in Table~\ref{table:1}.

 \begin{table}[h]
\caption{Interaction parameters of the SLJ potential. Amphiphilic, solvophobic and solvent are represented by A, H and S, respectively.}
 \begin{tabular}{c c c c c c c c }
 \hline
$ij$ \ \ \ \ & AA\ \ \ \ &HH\ \ \ \ &SS\ \ \ \ &AH\ \ \ \ &HS\ \ \ \ &AS\\
 \hline
 $\overline{\epsilon}_{ij}\ \ \ \ \ $&1\ \ \ \ &1\ \ \ \ &1\ \ \ \ &1\ \ \ \ &1\ \ \ \ & 1.4, 1.7, 1.8, 2.0\\
$\overline{r}_{{\rm c},ij}\ \ \ \ \ $&2.5\ \ \ \ &2.5\ \ \ \ &2.5\ \ \ \ &2.5\ \ \ \ &$2^{1/6}$\ \ \ \ &2.5\\
\hline
\end{tabular}
\label{table:1}
 \end{table}

Molecular dynamic simulations were performed in an NPT ensemble using the Nose-Hoover thermostat for different temperatures at a constant pressure $\overline{P}=0.002$. The trajectories were generated using the Velocity-Verlet algorithm with a time-step $\Delta \overline{t}= 0.004$. The ratio of the number of polymer beads to the solvent beads was maintained at 0.04 for all simulations. Simulations of $N=200$ chain were performed at four different interactions; $\overline{\epsilon}_{\rm AS}=1.4, 1.7, 1.8, 2.0$. For each of these values, the temperature was varied from $\overline{T}=0.5$ to 0.7 with an interval of 0.05. Simulations were also performed for a $N=400$ chain for $\overline{\epsilon}_{\rm AS}=1.7$ at the temperatures ranging from $\overline{T}=0.5$ to 0.8 with an interval of 0.05. The 200 and 400 bead systems were equilibrated for $1 \times10^{8}$ steps, and the data was sampled  after every $4 \times 10^{6}$ and $2 \times 10^{7}$ steps, respectively. Four different initial configurations were used for averaging. All simulations were performed using open source molecular dynamics code LAMMPS.\cite{Plimpton1995}

The simulation data were used for the calculation of different structural quantities. We calculated the radius of gyration, $R_{\rm g}$, of the polymer to monitor  the swelling of the polymer chain. We define a dimensionless radius of gyration $\overline{R}_{\rm g}=R_{\rm g}/\sigma$ given by the following expression
\begin{equation}\label{eq:rg}
\overline{R}_{\rm g}=\sqrt{\frac{1}{N}\sum_{i=1}^{N}(\overline{r}_{ i}-\overline{r}_{\rm  cm})^{2}},
\end{equation}
where  $N=200$ and $400$, $\overline{r}_{\rm cm}$ and $\overline{r}_{i}$ are the dimensionless coordinates of the centre of mass of the polymer chain and the $i$-th bead, respectively.

The number of solvent beads, $N_{\rm s}$,  in the first solvation shell of the polymer was calculated to determine the bound solvent content. 
The solvent beads which were within a distance of $\overline{r}=r/\sigma=1.5$ from any of the polymeric beads were regarded to be part of the first solvation shell.

The effective interaction between the polymer beads was calculated using the potential of mean force $U_{\rm AH}$, which was calculated from the  radial distribution function of  the amphiphilic and solvophobic bead pairs  using the following expression:\cite{Chandler1987}
\begin{equation}\label{eq:pmf}
\frac{\overline{U}_{\rm AH}(\overline{r})}{\overline{T}}=-\ln{g_{\rm AH}(\overline{r})}.
\end{equation}
where $\overline{U}_{\rm AH}=U_{\rm AH}/\epsilon_{\rm ss}$ is the dimensionless potential of mean force.

\subsection{Solvophobic Potential by Kolomeisky and Widom}
\label{sec:widom}

Our simulations indicate that the LCST is dependent on the entropy loss of the bound solvent, and the mean interaction energy difference between the bulk and the bound solvent (see Sec.~\ref{subsec:simulation}). To obtain further insights related to the nature of the transition, scaling behavior, and the effect of chain flexibility, a theoretical approach is adopted where we  consider the hydrophobic potential proposed by Kolomeisky and Widom (KW).\cite{Kolomeisky1999} The KW model is spherically symmetric in nature and does not relate to a specific solvent or solute, which is in correspondence with the modeling framework employed in our simulations. The KW model belongs to a class of implicit solvent models \cite{Lee1996,Moelbert2003a} which incorporate different interaction energies depending on the proximity of the solvent to a solute molecule. These models have been used for studying the solubility of small solutes in water. The KW model is one of the simplest among these models as it has only two solvent interaction energies; one for the bound state and another for the bulk state. These interaction energies are analogous to the monomer-solvent interaction energy ($\epsilon_{\rm AS}$), and solvent-solvent interaction energy ($\epsilon_{\rm SS}$) in our simulation model, respectively. Hence it can be seen that the model contains those contributions  which have been emphasized in our simulation results.

\begin{figure*}
\centering
\includegraphics[scale=0.3]{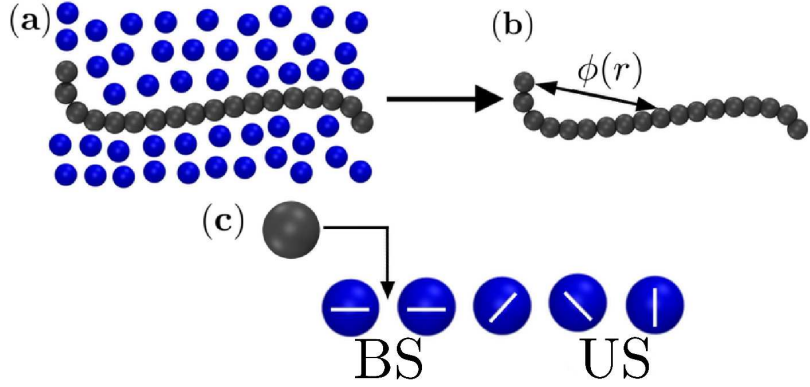}
\caption{Schematic representation for the theoretical model. Black and blue beads represent monomer and solvent molecules, respectively. (a) Single polymer chain in explicit solvent, (b) single polymer chain in implicit solvent where the effect of solvent is incorporated in the monomer-monomer interaction potential $\phi(r)$, and (c) $\phi(r)$ modeled by the solvophobic potential given by KW.\cite{Kolomeisky1999}  The potential assumes a one-dimensional solvent lattice where each solvent molecule has $q$  states. Neighboring solvent molecules exist in a bounded state (BS)  when both are in the  state \enquote{1} and in unbounded state (US) state otherwise. Solute can occupy interstitial sites between bounded (BS) solvent molecules. }
\label{fig:lattice}
\end{figure*}
\begin{figure}[h!]
\begin{center}    
\includegraphics[scale=0.35]{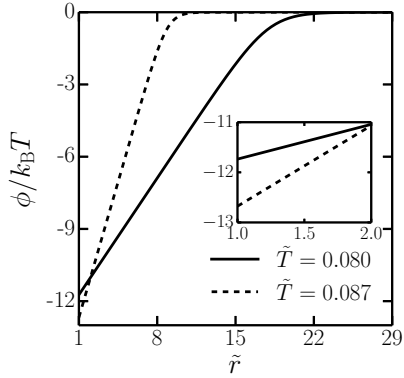}
\end{center}
\caption{Variation of $\phi/k_{\rm B}T$ as function of $\tilde{r}=r/\sigma$ for different 
$\tilde{T}$ when $q=5 \times 10^5$. 
The inset shows the variation at low $\tilde{r}$ values.} 
\label{fig:wr}
\end{figure}

In the KW-model, solvent molecules form a one-dimensional lattice with a nearest neighbor interaction, and each solvent molecule can exist in $q$ different  states denoted by $1, 2, \cdots q$
as shown in Fig.~\ref{fig:lattice}. The interaction energy between the neighboring solvent molecules is $w$ when both of them are in the state \enquote{1}, and $u$ otherwise with  $u>w$. Here the former and the latter cases are termed as the bound state (BS) state and the unbound state (US) state, respectively. A solvent molecule in the BS state can exist only in one state, whereas that in the US state in $q-1$ states. Hence the entropy of a solvent molecule in the BS state vanishes, whereas that in the US state is given by $k_{\rm B}\ln{(q-1)}$. In other words, the BS state is energetically favorable ($w<u$), while the US state is entropically favorable. The energetic ($\Delta U$) and entropic ($\Delta S$) differences between the BS and US states are $w-u$ and $-k_{\rm B}\ln{(q-1)}$, respectively. The competition between the BS and the US states can be conveniently described by a dimensionless parameter $x$ defined as   
\begin{equation}\label{eq:8}
x=e^{(\Delta U-T\Delta S)/k_{\rm B}T}=\frac{q-1}{c},
\end{equation}
where 
\begin{equation}\label{eq:9}
c=e^{(u-w)/k_{\rm B}T}.
\end{equation}
Since the number of  states $q$ is constant in the KW model, $x$ is a monotonically increasing function of the temperature.  Solute molecules are allowed to occupy only the interstitial sites between the solvent molecules in the BS state. Based on these assumptions, KW obtained the solvent mediated attraction potential $\phi(r)$ between two solute molecules (implicit solvent) for $r>\sigma,$\cite{Kolomeisky1999}
\begin{equation}\label{eq:widom_potential}
\phi(r) = -k_{\rm B}T\ln{\left[1+\left(\frac{1+Q}{1-Q}\right)
\left(\frac{1-S}{1+S}\right)^{(r-\sigma)/\sigma}
\right]},
\end{equation} 
where
\begin{equation}\label{eq:6}
S=\left[1-\frac{4 x}{(1+x)^2}\left(1-\frac{1}{c}\right)\right]^{1/2},
\end{equation}
\begin{equation}\label{eq:7}
Q=\frac{\rm \sgn{({\it x}-1)}}{\left[1+4 x/(x-1)^2 c\right]^{1/2}}, 
\end{equation}
and $\sgn(z)$ is the sign function.  Fig.~\ref{fig:wr} shows the variation of $\phi/k_{\rm B}T$ as a function of $\tilde{r}=r/\sigma$ for two different temperatures. The range of the  solvent mediated interaction becomes shorter  when the temperature is increased. From the inset of Fig.~\ref{fig:wr}, it can be  seen that the attraction becomes  stronger for higher temperature when $\tilde{r}$ is small. Such a behavior indicates that the monomers tend to aggregate as the temperature is increased. 
The large value of $q$ used by KW was justified in order to match the temperature dependence of the 
solubility of non-polar solutes in water.\cite{Kolomeisky1999} From Eq.~(\ref{eq:widom_potential}) it can be seen that the monomer-monomer potential is temperature dependent due to the implicit nature of solvent. An important point to note is that this temperature dependence is not ad-hoc but a consequence of the underlying solvent model  summarized in Fig.~\ref{fig:lattice}.

\section{Results and Discussions}
\label{sec:results}
\subsection{Simulation}
\label{subsec:simulation}

\begin{figure}[h]
\begin{center}    
\subfigure{\label{fig:a}\includegraphics[scale=0.35]{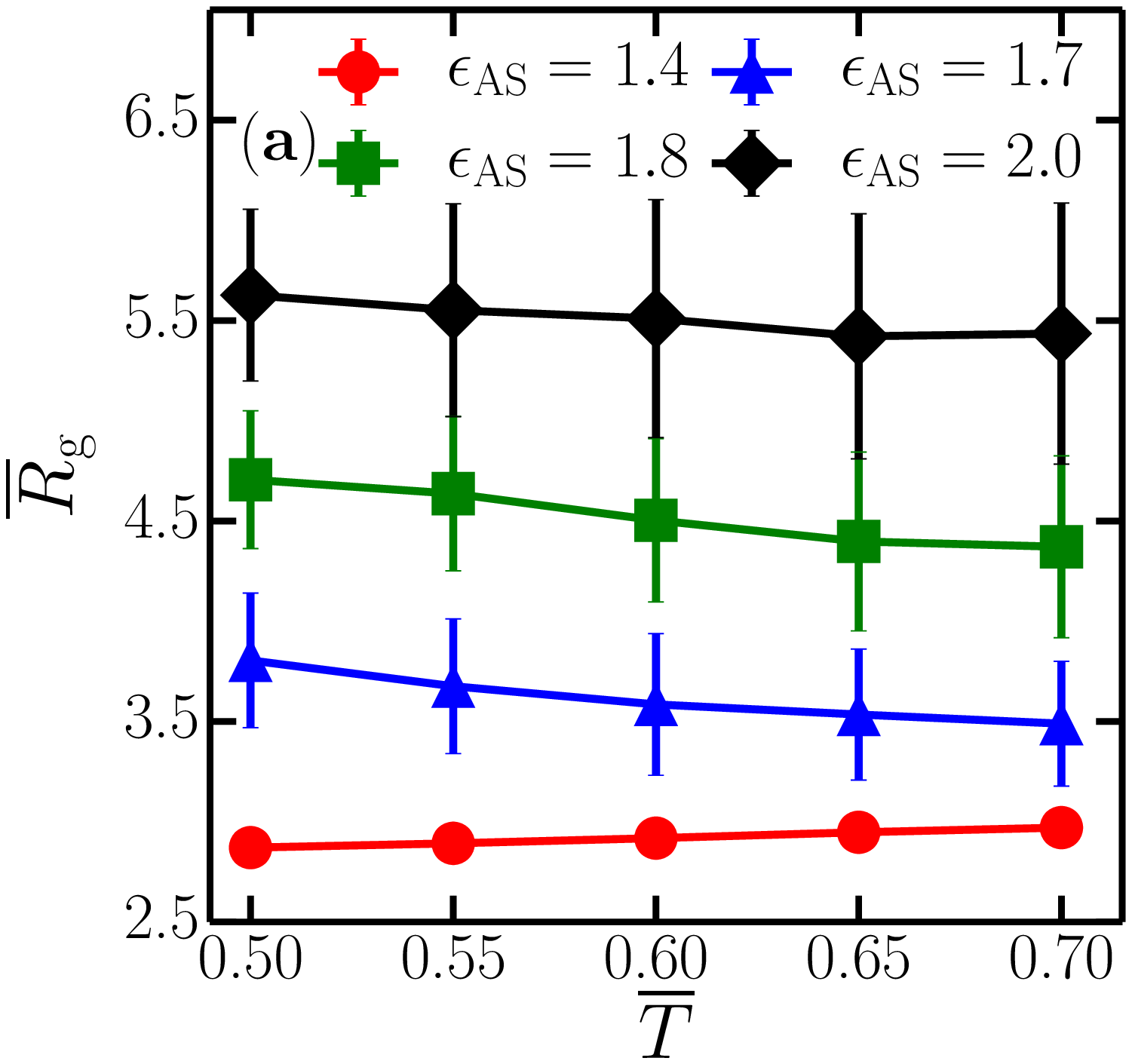}}
\subfigure{\label{fig:b}\includegraphics[scale=0.35]{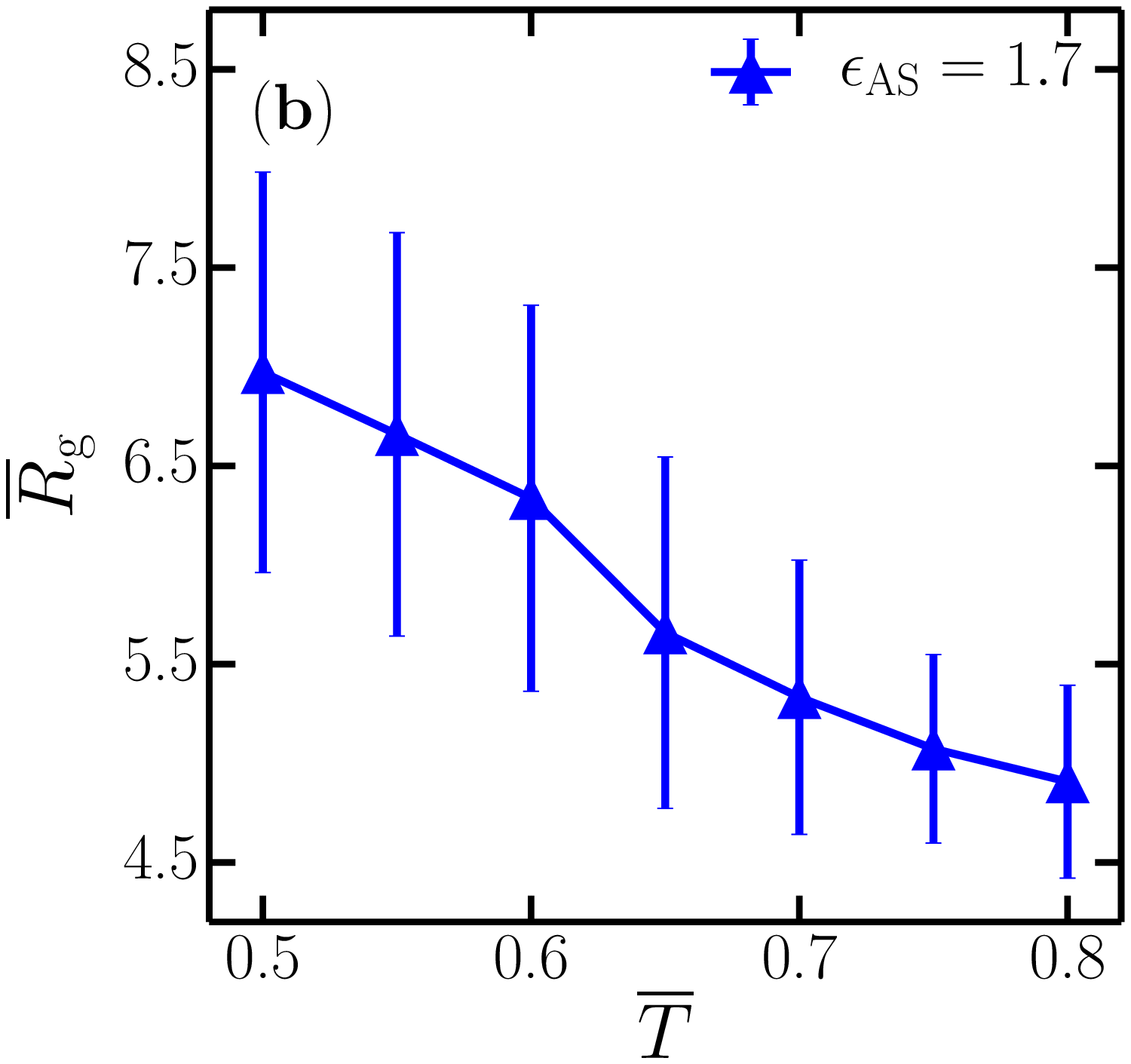}}
\end{center}
\caption{Variation of $\overline{R}_{\rm g}$  of polymer chain with $\overline{T}$ for (a) 200 bead polymer at different values of $\overline{\epsilon}_{\rm AS}$, (b) $N=400$ chain at $\overline{\epsilon}_{\rm AS}=1.7$.}
\label{fig:rg}
\end{figure}

To examine the structural change of the polymer chain with the temperature, we plot in Fig.~\ref{fig:a} the variation of $\overline{R}_{\rm g}$ with $\overline{T}$ for different $\overline{\epsilon}_{\rm AS}$ for the $N=200$ chain. The measured $\overline{R}_{\rm g}$ is distinctly larger when $\overline{\epsilon}_{\rm AS}$ is increased, implying  swelling as a result of stronger association between the amphiphilic bead and the solvent. We observe that there are three different behaviors according to the value of $\overline{\epsilon}_{\rm AS}$; (i) when $\overline{\epsilon}_{\rm AS}=1.4$, the polymer chain remains in a globular state and its  $\overline{R}_{\rm g}$ is almost independent of $\overline{T}$, (ii) when $\overline{\epsilon}_{\rm AS}=2.0$, the polymer chain is in the coiled state at all the temperatures, and (iii) when $\overline{\epsilon}_{\rm AS}=1.7$ and $1.8$, $\overline{R}_{\rm g}$ decreases with the temperature, which is similar to the LCST behavior in thermoresponsive polymers. 

\begin{figure*}[h]
\centering
\includegraphics[scale=0.48]{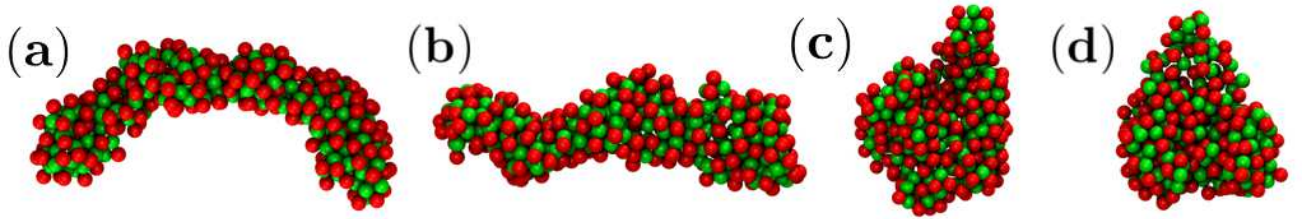}
\caption{Representative snapshots of equilibrated $N=400$ chain at $\epsilon_{\rm AS}=1.7$ for different temperatures. Green and red beads represent solvophobic and amphiphilic units, respectively. (a) $\tilde{T}=0.55$, (b) $\tilde{T}=0.6$, (c) $\tilde{T}=0.7$ and (d) $\tilde{T}=0.8$. Solvent beads are not included for clarity.}
\label{fig:snap}
\end{figure*}

To further probe the decrease in $\overline{R}_{\rm g}$ with the temperature, we focus on the $N=400$ chain for $\overline{\epsilon}_{\rm AS}=1.7$. From Fig.~\ref{fig:b}, we see that the size change with the temperature for the $N=400$ chain is more prominent compared to that of the $N=200$ chain, which is due to difference in scaling of  $\overline{R}_{\rm g}$ with $N$ for the coiled and globular states.\cite{Grosberg1997} In Fig.~\ref{fig:b}, $\overline{R}_{\rm g}$ decreases by 23\% as the temperature is increased from $\overline{T}=0.55$ to 0.75. Such trends have been observed in atomistic simulations of PNiPAM-water system.\cite{Du2010a,Alaghemandi2012a, Deshmukh2012b} Unlike the first-order LCST transition behavior observed in experimental studies of thermoresponsive polymer solutions,\cite{Wu1998, Wang1999, Cao2005}  the transition observed in Fig.~\ref{fig:b} shows a gradual change in $\overline{R}_{\rm g}$ with $\overline{T}$. It is known that the coil-to-globule transition is a first-order transition only in the case of rigid and semirigid chains.\cite{Grosberg1992, Graziano2000, Baysal2003} Hence the continuous transition in our simulation is not surprising because the polymer chain used in the simulations, is fully flexible due to the absence of the angular and dihedral interactions. This is further supported by the results of our theoretical model, where we observe that the coil-to-globule transition deviates from a first order behavior with increase in chain flexibility (see Sec~\ref{sec:khok}). Based on the above observations, we consider that the behavior  observed for $\overline{\epsilon}_{\rm AS}=1.7$ is akin to the LCST phenomenon in thermoresponsive polymers. Figure ~\ref{fig:snap} shows representative snapshots of $N=400$ chain at $\epsilon_{\rm AS}=1.7$ for different temperatures, we can see that the polymer chain is in the coil-like state below $\overline{T}=0.6$, while it is is the globule-like 
state above $\overline{T}=0.7$.

\begin{figure}
\centering    
\subfigure{\label{fig:wa}\includegraphics[scale=0.35]{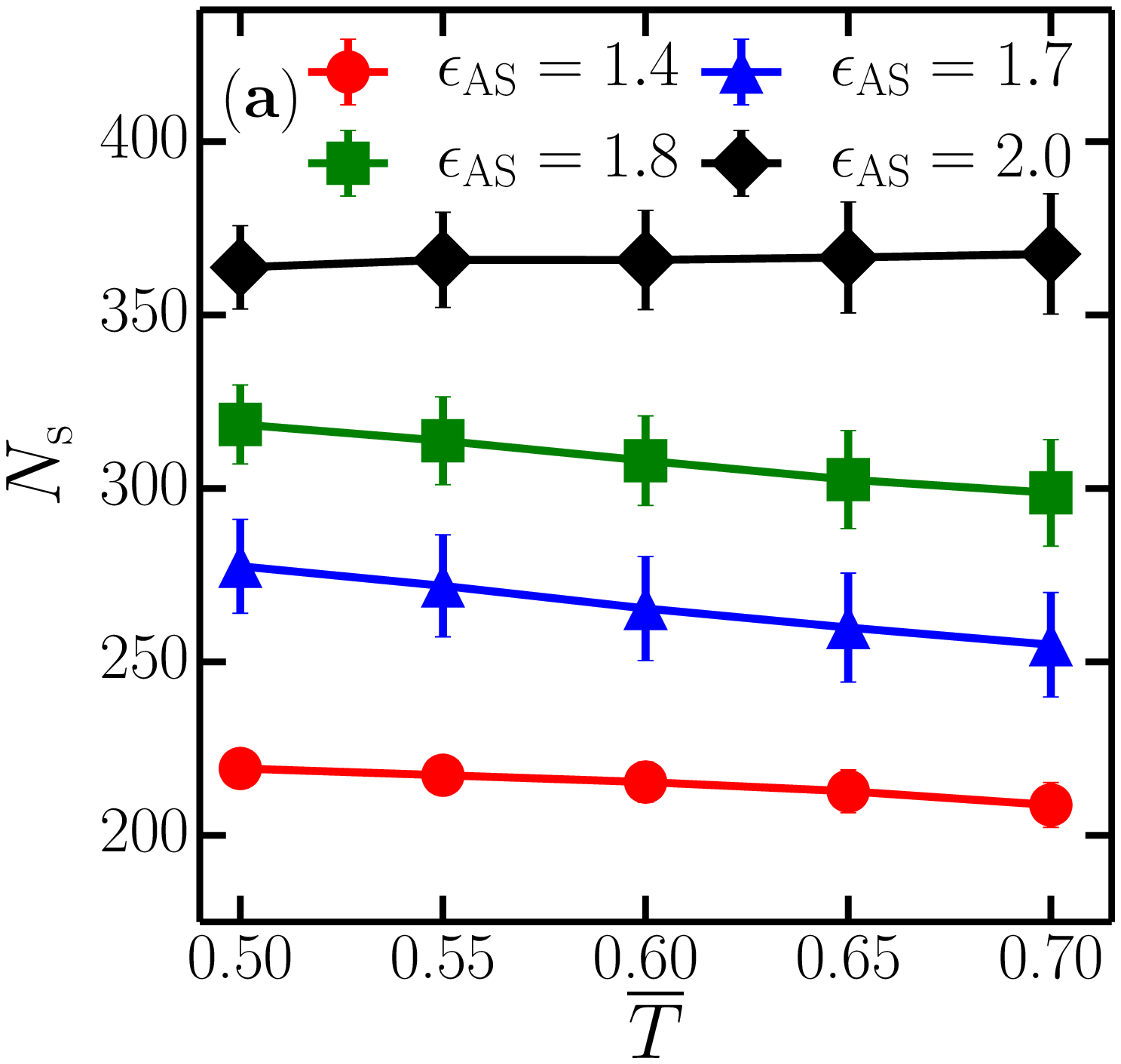}}
\subfigure{\label{fig:wb}\includegraphics[scale=0.35]{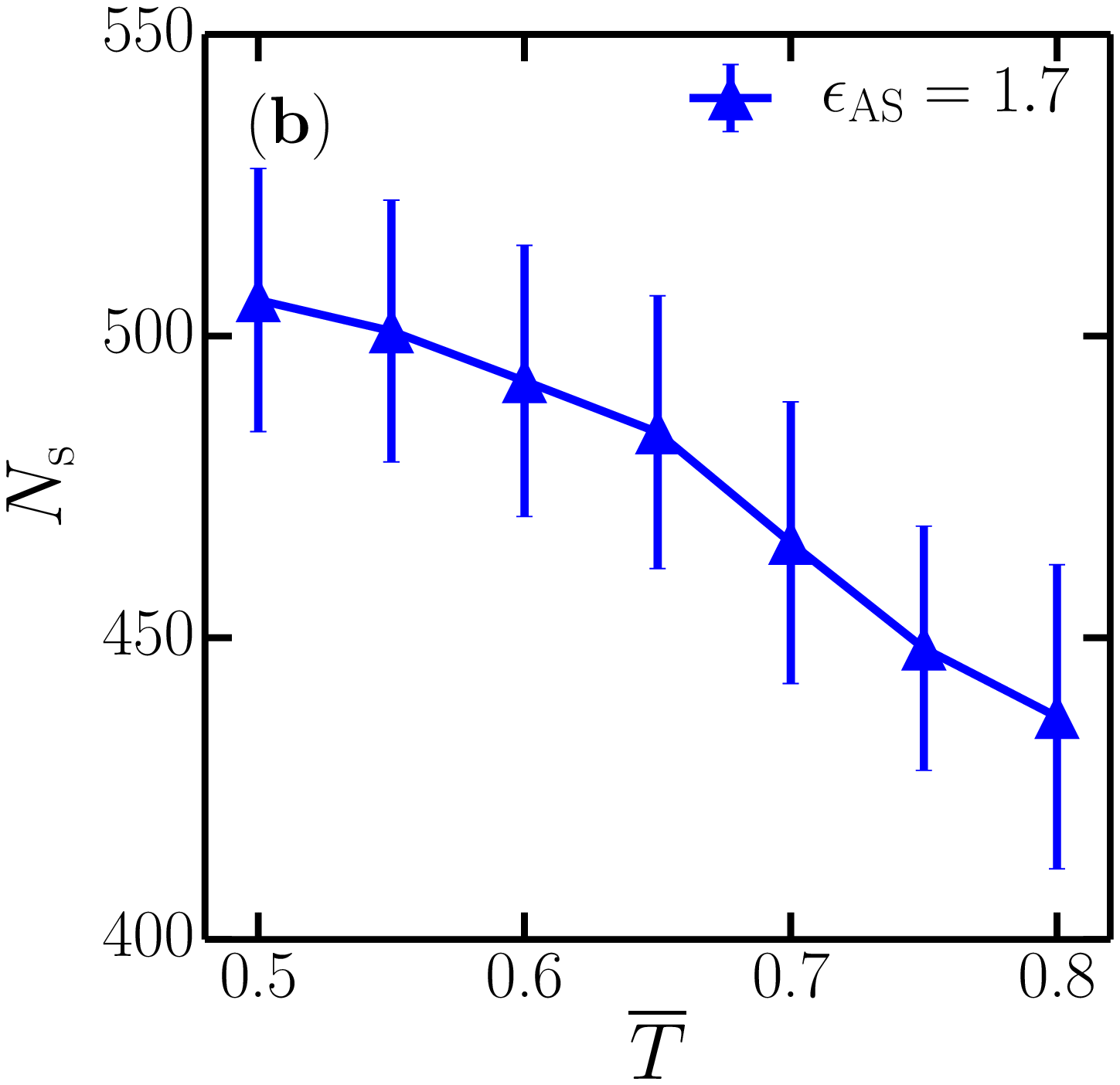}}
\caption{(a) Variation of  $N_{\rm s}$ with $\overline{T}$ at different values for (a) $N=200$ chain at different values of $\overline{\epsilon}_{\rm AS}$, (b) $N=400$ chain at $\overline{\epsilon}_{\rm AS}$=1.7.}
\label{fig:water}
\end{figure}

Hereafter we will be referring to $\overline{\epsilon}_{\rm AS}=1.4$, $\overline{\epsilon}_{\rm AS}=1.7$ and $1.8$, and $\overline{\epsilon}_{\rm AS}=2.0$ as low, intermediate and high values, respectively. To further understand the behavior in these three states, we examine the variation of the number of bound solvent beads, $N_{\rm s}$,  with the temperature.  In Fig.~\ref{fig:wa}, the number of bound solvent is found to be almost independent of $\overline{T}$ for low and high values of $\overline{\epsilon}_{\rm AS}$, whereas it decreases by increasing the temperature for intermediate values. In Fig.~\ref{fig:wb}, we plot $N_{\rm s}$ for the $N=400$ chain as a function of $\overline{T}$, when $\epsilon_{\rm AS}=1.7$. The decrease of $N_{\rm s}$ with the temperature is more prominent as compared with that of the $N=200$ chain case. Moreover, $N_{\rm s}$ markedly decreases around $\overline{T}=0.65$ which coincides with the temperature around which $\overline{R}_{\rm g}$ also decreases (see Fig.~\ref{fig:b}). It should be stressed that the change in $\overline{R}_{\rm g}$ and $N_{\rm s}$ with the temperature is observed in our model with minimal interactions as in Eq.~(\ref{eq:energy_simu}). It is important to note that  even when the interaction parameters are independent of the temperature, the  temperature dependence of  $\overline{R}_{\rm g}$  is induced by the bound solvent number  $N_{\rm s}$ for a range of values of  $\overline{\epsilon}_{\rm AS}$.

\begin{figure}
  \centering
\subfigure{\label{fig:pmf_lcst}\includegraphics[scale=0.35]{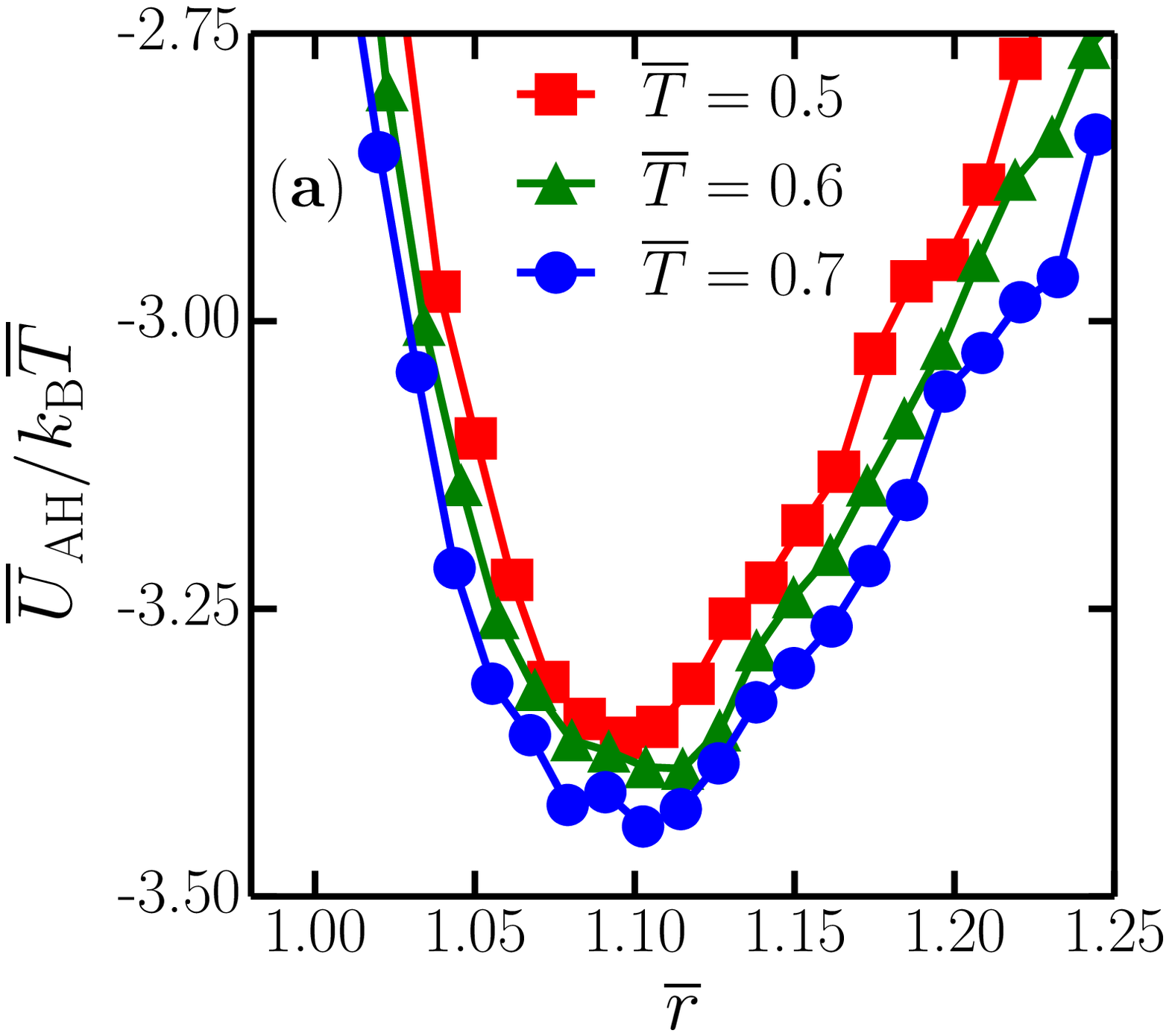}}
\subfigure{\label{fig:pmf_collapsed}\includegraphics[scale=0.35]{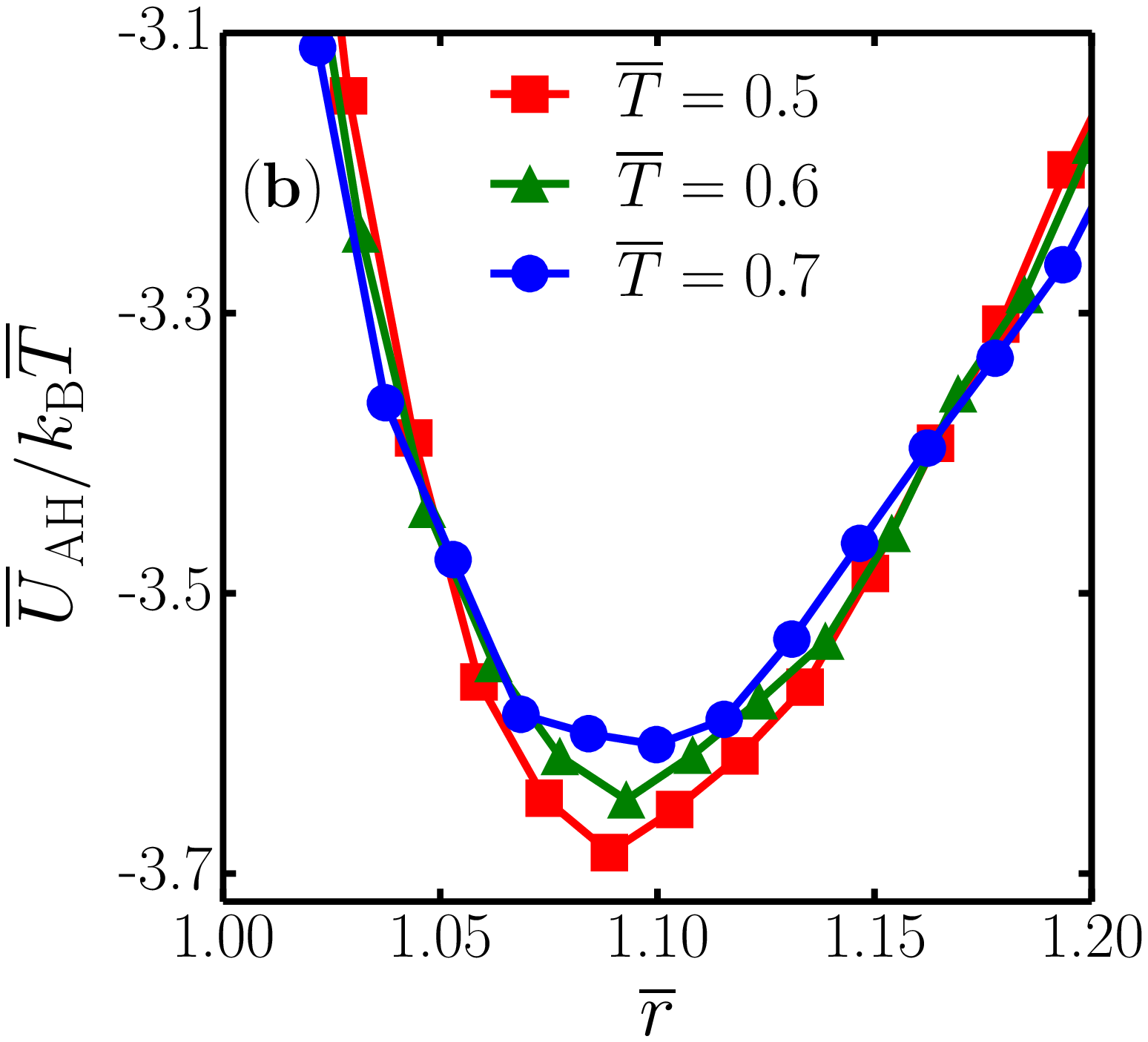}}
\caption{Variation of $\overline{U}_{\rm AH}/k_{\rm B}\overline{T}$ with $\overline{r}$ at different temperatures for (a) $N=400$ chain at $\overline{\epsilon}_{\rm AS}=1.7$  (b) $N=200$ chain at $\overline{\epsilon}_{\rm AS}=1.4$.} 
\end{figure}

The above trends in the three different states can be rationalized  by considering the different energy, and entropic contributions in the model system. 
The attraction between the amphiphilic monomer and the solvent is energetically favorable since
it is stronger than the solvent-solvent interaction ($\epsilon_{\rm AS} > \epsilon_{\rm SS}$), leading to a coil-like polymer conformation with 
many bound solvent beads. Whereas from the viewpoint of entropy, the solvent prefers to be in the bulk state rather than the bound state.  Another important contribution is the solvation of solvophobic beads, which is unfavorable as  the interaction between the solvophobic bead and the solvent is repulsive.  
The interplay of these contributions determines the different states of the polymer chain. \\
\indent Concerning intermediate $\overline{\epsilon}_{\rm AS}$ values, the large values of $\overline{R}_{\rm g}$ in the coil-like state
(see Fig.~\ref{fig:b}) and large $N_{\rm s}$ at low temperatures (see Fig.~\ref{fig:wb})  indicate 
that the attraction between the amphiphilic monomer and the solvent dominates the entropy 
loss of the solvent. 
Hence it is favorable for the solvent to solvate the solvophobic beads despite the repulsive 
interaction between them. With increase in the temperature, the gain in the entropy  due to unbinding of the  solvent  leads to a reduction in the bound solvent content, $N_{\rm s}$. In the transition region ($0.6 \leq T \leq 0.7$), there is marked decrease of the bound solvent 
because the entropy  gain of the solvent dominates over the amphiphilic-solvent attraction.
This change in the dominant contribution makes the solvation of solvophobic beads unfavorable, leading to an attraction between the polymeric beads. 
Such an attraction becomes stronger with the temperature, and drives the transition from 
the coil-like state to the globule-like state. 
In Fig~\ref{fig:pmf_lcst}, we plot the potential of mean force, $\overline{U}_{\rm AH}$, 
between the amphiphilic and the solvophobic beads as a function of the distance $\overline{r}$.
Here we see that the attraction between the polymeric beads increases with the temperature. At low temperature, the attraction between the polymeric beads is screened due to the presence of solvent beads. At high temperatures, the bound solvent beads unbind due to entropy gain which reduces the screening effect, leading to the collapse of the polymer. As shown in Fig.~\ref{fig:pmf_collapsed}, the increase in the  attraction between polymeric beads with temperature is not observed for low $\overline{\epsilon}_{\rm AS}$ values.\\
\indent For low $\overline{\epsilon}_{\rm AS}$ values, both $\overline{R}_{\rm g}$ and $N_{\rm s}$ are small.
This means that  at these interaction strengths the  entropic  gain  of the free solvent beads dominates over the interaction between 
the amphiphilic monomer and the solvent for all the temperatures.
Hence the solvation of solvophobic beads is unfavorable  in the 
entire temperature range.  On the other hand, for high $\overline{\epsilon}_{\rm AS}$ values, both $\overline{R}_{\rm g}$ and $N_{\rm s}$ are large.
This indicates that the attractive interaction is stronger than the entropy loss which leads to the binding of the solvent to the polymer. Hence, the solvation of the solvophobic beads is favored at all temperatures  despite the repulsive interaction between the solvophobic monomer and the solvent.

We find from our simulations that the transition from the coil-like state to the globule-like state 
depends on how much the solvophobic beads can be solvated. 
The solvation depends on the interplay between the entropy loss of the bound solvent,  and the energetic
difference between the bound and the bulk solvent. 
The bound solvent is energetically favored as the interaction between the amphiphilic monomer
and the solvent is stronger than the solvent-solvent interaction.
Therefore two kinds of beads having opposite interactions with the solvent are necessary to exhibit 
the LCST as long as the interaction parameters are independent of the temperature as in our case. 
Additionally in our simulation, we did not incorporate any specific chemical or structural details pertaining to 
the solvent or the polymer chain.
 Therefore we have demonstrated that the LCST behavior can be exhibited by a coarse-grained description of 
the polymer and the solvent even  when the interaction parameters are independent of the temperature.
\subsection{ Theoretical description  of the  coil-to-globule transition}
\label{sec:khok}
 To study the coil-to-globule transition in the theoretical framework, we adopt the phenomenological free energy expression for a polymer chain in an implicit solvent, which has been  given by Grosberg and Kuznetsov.\cite{Grosberg1992}   We choose the KW potential (see Sec~\ref{sec:widom}) to model the effective monomer-monomer interaction. The free energy expression for the system is as follows:

\begin{equation}
\frac{F}{k_{\rm B}T}=\alpha^{2}+\frac{1}{\alpha^{2}} +\frac{\sqrt{N}}{\alpha^{3}\sigma^{3}}B+
\frac{1}{\alpha^{6}\sigma^{6}}C,
\label{eq:4}
\end{equation}
where $\alpha=R_{\rm g}/(\sqrt{N}\sigma)$ characterizes the extent of the swelling,
$N$ the degree of polymerization, $\sigma$ the diameter of the monomeric unit, 
$B$ and $C$ are the second and the third virial coefficients, respectively. 
The first and the second terms in the above free energy are the entropic contributions with the 
the coiled and the globular states given by $\alpha>1$ and  $\alpha<1$, respectively. 
The third and the fourth terms represent the energy contributions from the two-body and the three-body interactions, respectively. 
In general, the second virial coefficient $B$ is given by 
\begin{equation}
B=2\pi\int_{0}^{\infty} {\rm d}r \, r^2 \left(1-e^{-\Phi(r)/k_{\rm B}T}\right), 
\label{eq:second_virial}
\end{equation}
where $\Phi(r)$ is the monomer-monomer interaction potential.
Here we assume that it has the following form:
\begin{equation}\label{eq:interaction}
\Phi(r) = \left\{
\begin{array}{ll}
\infty &: r<\sigma \\
\phi(r) &: r>\sigma 
\end{array}
\right.,
\end{equation} 
where $\Phi(r)$ for $r<\sigma$ corresponds to the (hard-core) excluded volume 
interaction, and $\phi(r) $ for $r>\sigma$ (see Eq.~(\ref{eq:widom_potential})) is the KW solvophobic potential.
Substituting Eqs.~(\ref{eq:widom_potential}) and ~(\ref{eq:interaction}) into Eq.~(\ref{eq:second_virial}), we obtain the second 
virial coefficient $B$ as 
\begin{widetext}
\begin{equation}
B=\frac{2\pi \sigma^{3}}{3}\left[1+ 3\left(\frac{1+Q}{1-Q}\right)\left( \frac{1}{\ln{L}} - 
\frac{2}{\left(\ln{L}\right)^{2}} +\frac{2}{\left(\ln{L}\right)^{3}} \right)\right],
\label{eq:11_g}
\end{equation}
\end{widetext}
where $L=(1-S)/(1+S)$.

\begin{figure}
\begin{center}    
\includegraphics[scale=0.35]{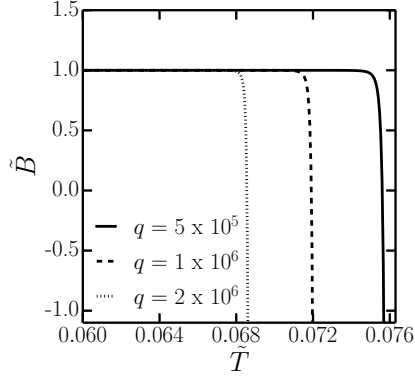}
\end{center}
\caption{Variation of $\tilde{B}$ with $\tilde{T}$ for different $q$ values.}
\label{fig:br}
\end{figure}

Hereafter we use the dimensionless quantities such as 
the temperature $\tilde{T}=k_{\rm B}T/(u-w)$, 
the distance $\tilde{r}=r/\sigma$,
the solvent mediated interaction potential $\tilde{\phi}=\phi/(u-w)$, 
and the second virial coefficient $\tilde{B}=3B/2\pi \sigma^{3}$. 
In Fig.~\ref{fig:br}, we plot $\tilde{B}$ as a function of $\tilde{T}$ for different
$q$-values.  
We observe that $\tilde{B}$ remains almost unity for low temperatures and then decreases 
rapidly for higher temperatures.
This means that the monomer-monomer interaction is repulsive for lower temperatures, 
while it is attractive for higher temperatures. 
Moreover, the temperature corresponding to the sharp drop from the positive 
(repulsive) to the negative (attractive) $\tilde{B}$-values decreases when $q$ is increased  for a fixed  $u-w$ value. The significance of this observation will be  discussed later. 

\begin{figure}
\centering
\includegraphics[scale=0.35]{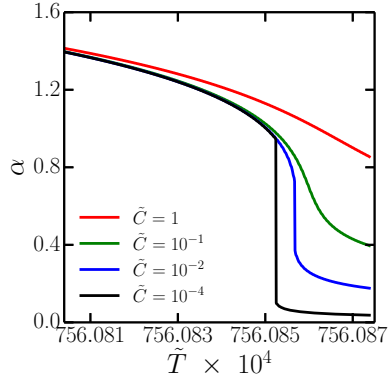}
\caption{Variation of $\alpha$ with $\tilde{T}$ for different values of $\tilde{C}$
when $q=5 \times 10^5$ and $N=10^5$.}
\label{fig:ctog}
\end{figure}

In order to find the equilibrium polymer conformation, we minimize Eq.~(\ref{eq:4}) 
with respect to $\alpha$, and obtain the equation
\begin{equation}\label{eq:minimisation}
\alpha^{5}-\alpha-\pi\sqrt{N}\tilde{B} - \frac{\tilde{C}}{\alpha^{3}}=0,
\end{equation}
where $\tilde{C}=3C/\sigma^{6}$ is the non-dimensional third virial coefficient which is related to the rigidity of the polymer chain and has contributions from the three-body interactions such as angular interactions.
We numerically solve the above equation to obtain $\alpha$ for fixed values of $q$, $N$ 
and $\tilde{C}$. 
In Fig.~\ref{fig:ctog}, we show the variation of the swelling parameter 
$\alpha$ with the temperature $\tilde{T}$ for different values of the 
third virial coefficient $\tilde{C}$ when $q=5 \times 10^{5}$.  
We see that the polymer chain undergoes a transition from the coiled state to 
the globular state with increase in the temperature.

For low temperatures, the polymer chain is in the coiled state and hence large $\alpha$.
In this case, the first and the third terms in Eq.~(\ref{eq:minimisation}) are dominant
because $N$ is also large. 
Then Eq.~(\ref{eq:minimisation}) reduces to
\begin{equation}\label{eq:11_h_1}
\alpha^{5}-\pi\sqrt{N}\tilde{B} \approx 0,
\end{equation}
and we obtain $R_{\rm g} \sim N^{3/5}$ corresponding to the scaling 
of a polymer chain in a good solvent.
When the temperature is high, on the other hand, the polymer is in the globular 
state and $\alpha$ becomes small.     
Then the third and the fourth terms dominate in Eq.~(\ref{eq:minimisation});
\begin{equation}\label{eq:11_h}
\pi\sqrt{N}\tilde{B} + \frac{\tilde{C}}{\alpha^{3}} \approx 0.
\end{equation}
This gives the scaling $R_{\rm g} \sim  N^{1/3}$ corresponding to a polymer 
chain in a poor solvent.

In order to understand this coil-to-globule transition in terms of the dominant interactions, 
let us consider the free energy difference of a solvent molecule between the BS and US states;

\begin{equation}\label{eq:deltaf}
\Delta F = F_{\rm BS}-F_{\rm US}=U_{\rm BS}-U_{\rm US} - T(S_{\rm BS}-S_{\rm US}),
\end{equation}
where $U_{\rm BS}$ ($U_{\rm US}$) and $S_{\rm BS}$ ($S_{\rm US}$) are the 
energy and the entropy of the BS (US) state, respectively.
Using the model parameters defined in Sec.~\ref{sec:widom}, the above quantity can be expressed as 
\begin{equation}
\Delta F=-(u-w) +k_{\rm B}T \ln{(q-1})=k_{\rm B}T\ln x,
\label{eq:DeltaF}
\end{equation}
where $x$ is defined before in Eq.~(\ref{eq:8}). From Eq.~(\ref{eq:DeltaF}) and  since $u>w$, we find $\Delta F<0$  for $x<1$ (lower temperatures), leading to the BS state being more favorable. 
Given the implicit nature of the solvent in the theoretical treatment, the large value 
of $\alpha$ for lower temperatures in Fig.~\ref{fig:ctog} is an indication of the large 
amount of the bound solvent.  
For $x>1$ (higher temperatures), on the other hand, the US state is more favorable and 
the amount of the unbound solvent increases. 
In this case, the attraction between the monomeric units is induced.
This is seen in Fig.~\ref{fig:br} for  the larger negative value of the second virial coefficient
$\tilde{B}$. 
These solvent induced interactions drive the transition from the coiled state to the 
globular state. 
The phase transition temperature $T^{\ast}$ determined by the condition $\Delta F=0$ 
is given by
\begin{equation}\label{eq:critical_temp}
T^{\ast}=\frac{u-w}{k_{\rm B}\ln{(q-1)}}.
\end{equation}
As observed in Sec.~\ref{sec:widom}, $T^{\ast}$ decreases as $q$ is increased when $u-w$ is fixed.

We further observe in Fig.~\ref{fig:ctog} that the nature of the transition changes from a discontinuous transition to a continuous one by increase in the third virial coefficient $\tilde{C}$. It is known that the third virial coefficient is larger for flexible chains.\cite{Grosberg1992,Graziano2000} This shows that the nature of transition deviates from a first-order behavior with increase in the flexibility of the chain. Another  important point is that the above argument does not include any details regrading the structure of the solvent and/or polymer which indicates that the structural details of the bound solvent are not necessary for the thermoresponsive behavior to manifest.

\section{Summary and Conclusions}
\label{sec:sum}
In this paper, we have tried to understand the single chain coil-to-globule transition of a thermoresponsive polymer through simulation and theoretical approaches. In the simulations, the model comprises of a single polymer chain in an explicit solvent with temperature independent interaction parameters. The solvent is explicitly included to avoid any ad-hoc dependence of the interaction potentials on temperature. To obtain further insights, we have adopted a theoretical  framework with only those interactions that have been emphasized in our simulation studies. The theoretical model describes a single chain in an implicit solvent where the effect of solvent is included into the monomer-monomer interaction potential. For the interaction between the monomers, we have used the solvophobic potential proposed by Kolomeisky and Widom.\cite{Kolomeisky1999} The temperature dependence of the solvophobic potential is not ad-hoc in nature and arises due to the underlying solvent model.\\
\indent Our simulations indicate that the LCST is dependent on the competition between 
the two contributions, namely, the entropy loss of the bound solvent and the mean energy 
difference between the bound and the bulk solvent. This hypothesis is supported by our theoretical calculations.  
The former favors the globular state of the polymer chain, whereas the latter prefers the coiled state.
At low temperatures, solvent molecules bind to the polymer chain rather than to reside in the bulk because 
the bound state is energetically favorable, and the coiled state is obtained. 
At high temperatures, on the other hand, the entropy loss of the bound state is more dominant than 
its energetic gain and the solvent molecules tend to leave the polymer. Such a competition of the interactions induces a solvent driven attraction between the polymeric beads 
leading to the collapse of the polymer chain. These findings are supported by the experimental studies of Bischofberger and coworkers who showed that the LCST is dependent on the mean energy difference between the bound and bulk solvent.\cite{Bischofberger2014, Bischofberger2014a}\\
\indent Another common feature between the simulation and the theory is that both of them have spherically symmetric solvent and monomeric beads. This indicates that the structural arrangement of the solvent molecules around the polymer chain is not a necessary component to be considered explicitly  for the coil-to-globule transition. In other words, a coarse-grained description of the solvent is sufficient to reproduce the LCST behavior, this observation is in agreement with the experimental results of Bischofberger and coworkers,\cite{Bischofberger2014, Bischofberger2014a} where they show that a coarse-grained representation of the solvent is sufficient to explain the LCST and its variation with different alcohols.\\
\indent In our simulations studies, it is shown that the LCST is dependent on the amphiphilic-solvent attraction and solvation of the solvophobic beads. This indicates that in the case of temperature-independent interaction parameters, two different kinds of beads with opposite interaction with the solvent are required in the monomeric unit to exhibit LCST. In the  theoretical model, the solvent is implicit and the variation of solvent effects with temperature is included in the monomer-monomer interaction due to which one kind of bead is sufficient to exhibit LCST. A mixture of PNiPAM and water exhibits a first-order phase transition at 32$\degree$C.\cite{Wang1998, Wu1998} In our theoretical argument, we showed that the order of the transition changes from a first-order (discontinuous) to a second-order (continuous) with the increase in the chain flexibility (or the third virial coefficient). The reason why we observe only the continuous coil-to-globule transition in our simulation is because we did not take into account any angular interactions to deal with the chain flexibility. The effect of angular and dihedral interactions on the nature of the coil-to-globule transition needs to be explored further. \\
\indent The variation in behavior with temperature for different $\overline{\epsilon}_{\rm AS}$ values (see Fig~\ref{fig:a}) is analogous to different polymers in the family of poly(N,N-alkyl alkyl acrylamide). Low $\overline{\epsilon}_{\rm AS}$ values are similar to polymers with bulky side groups such as poly(N-butyl acrylamide) which are always insoluble in water. On the other hand, high $\overline{\epsilon}_{\rm AS}$ values are similar to polymers such as polyacrylamide which are always soluble in water. This indicates that our coarse-grained model is able to explain the LCST behavior for different systems in the poly(N,N-alkyl,alkyl acrylamide) family of polymers. Furthermore, the model can also be utilized to study the variation of LCST due to factors such as co-solvents and additives by examining their effect on the two dominant physical interactions namely, the entropy loss of the bound solvent and the mean energy difference between the bound and the bulk solvent. This aspect will be addressed in our future studies.
\begin{acknowledgments}
The computations were carried out at the High Performance Computing Facility at IIT Madras. S.K. acknowledges support from the Grant-in-Aid for Scientific Research on Innovative Areas ``\textit{Fluctuation and Structure}" (Grant No.\ 25103010) from the Ministry of Education, Culture, Sports, Science, and Technology of Japan, the Grant-in-Aid for Scientific Research (C) (Grant No.\ 15K05250) from the Japan Society for the Promotion of Science (JSPS), and the JSPS Core-to-Core Program ``\textit{International Research Network for Non-equilibrium Dynamics of Soft Matter}". G.S.B thanks Okamoto, Andelman, Koga and Seki for valuable discussions. 
\end{acknowledgments}

\end{document}